\documentclass[conference]{IEEEtran}

\usepackage{times}
\usepackage[utf8]{inputenc} 
\usepackage[T1]{fontenc}    
\usepackage{hyperref}       
\usepackage{url}            
\usepackage{booktabs}       
\usepackage{amsfonts}       
\usepackage{nicefrac}       
\usepackage{microtype}      
\usepackage{lipsum}		
\usepackage[numbers]{natbib}
\usepackage{doi}
\usepackage{amsmath}

\usepackage{graphicx}
\usepackage[retain-explicit-plus]{siunitx}
\usepackage{textcomp}
\usepackage{array}
\usepackage{datetime}
\newcolumntype{R}[1]{>{\raggedleft\let\newline\\\arraybackslash\hspace{0pt}}m{#1}}
\newcolumntype{L}[1]{>{\raggedright\let\newline\\\arraybackslash\hspace{0pt}}m{#1}}
\newcolumntype{C}[1]{>{\centering\let\newline\\\arraybackslash\hspace{0pt}}m{#1}}
\DeclareSIUnit\permille{\text{\textperthousand}}
\newboolean{IEEE_ACCESS_FORMAT}
\setboolean{IEEE_ACCESS_FORMAT}{false}

\def\BibTeX{{\rm B\kern-.05em{\sc i\kern-.025em b}\kern-.08em
    T\kern-.1667em\lower.7ex\hbox{E}\kern-.125emX}}
\begin{document}

\title{Leveraging Eclipse MOSAIC for Modeling and Analyzing Ride-Hailing Services}

\author{
\IEEEauthorblockN{Karl Schrab}
\IEEEauthorblockA{\textit{DCAITI} \\
\textit{Technical University of Berlin}\\
Berlin, Germany \\
karl.schrab@dcaiti.com \href{https://orcid.org/0000-0002-5083-595X}{\includegraphics[scale=0.07]{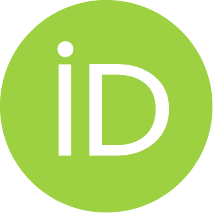}}}
\and
\IEEEauthorblockN{Moritz Schweppenh\"auser}
\IEEEauthorblockA{\textit{Smart Mobility} \\
\textit{Fraunhofer Institute FOKUS}\\
Berlin, Germany \\
moritz.schweppenhaeuser@fokus.fraunhofer.de \href{https://orcid.org/0009-0001-9252-2425}{\includegraphics[scale=0.07]{arxiv/orcid.pdf}}}
\and
\IEEEauthorblockN{Rober Protzmann}
\IEEEauthorblockA{\textit{Smart Mobility} \\
\textit{Fraunhofer Institute FOKUS}\\
Berlin, Germany \\
robert.protzmann@fokus.fraunhofer.de \href{https://orcid.org/0000-0002-5531-1936}{\includegraphics[scale=0.07]{arxiv/orcid.pdf}}}
\and
\IEEEauthorblockN{Kay Massow}
\IEEEauthorblockA{\textit{Smart Mobility} \\
\textit{Fraunhofer Institute FOKUS}\\
Berlin, Germany \\
kay.massow@fokus.fraunhofer.de \href{https://orcid.org/0000-0002-3760-5762}{\includegraphics[scale=0.07]{arxiv/orcid.pdf}}}
\and
\IEEEauthorblockN{Ilja Radusch}
\IEEEauthorblockA{\textit{DCAITI} \\
\textit{Technical University of Berlin}\\
Berlin, Germany \\
ilja.radusch@dcaiti.com}
}


\hypersetup{
pdftitle={Leveraging Eclipse MOSAIC for Modeling and Analyzing Ride-Hailing Services},
pdfauthor={Karl Schrab, Moritz Schweppenhäuser, Robert Protzmann, Kay Massow, Ilja Radusch},
pdfkeywords={Eclipse MOSAIC, Ride-Hailing Large-scale Traffic Simulation},
}

\maketitle

\begin{abstract}
Ride-hailing services enjoy a large popularity in the sector of individualized mobility.
Due to broad availability, ease of use, and competitive pricing strategies, these services have established themselves throughout the last decades.
With the increased popularity, ride-hailing providers aimed to consistently improve the efficiency of their services, leading to the inception of novel research questions.
Many of which can be effectively tackled using simulation.
In this paper, we present such a simulation-based approach using Eclipse MOSAIC in-hand with a large-scale traffic scenario of Berlin.
We analyze real-world logbook data including detailed shifts of drivers and discuss how to integrate them with the simulation scenario.
Moreover, we present extensions to MOSAIC required for the modeling of the ride-hailing services, utilizing the powerful Application Simulator.
Accordingly, as the primary result of this paper, we managed to extend the Eclipse MOSAIC framework to be able to answer research questions in the domain of ride-hailing and ride-sharing.
Additionally, in an initial exemplary study, we analyze the traffic and environmental impacts of different, yet basic, rebalancing strategies, finding non-negligible differences in mileages and pollutant emissions.
We, furthermore, applied our findings to the entire ride-hailing fleet in the city of Berlin for one year, showcasing the impacts different rebalancing strategies could have on environment and general traffic.
To our knowledge, the consideration of environmental factors on a city-wide scale is a novel contribution of this paper, not addressed in previous research.

\end{abstract}


\section{Introduction}
\label{sec:introduction}

\ifthenelse{\boolean{IEEE_ACCESS_FORMAT}}
{
\PARstart{R}{ide-hailing} and ride-pooling services have become an integral part of urban transportation systems, offering convenient and flexible transportation options for passengers.
}
{
Ride-hailing and ride-pooling services have become an integral part of urban transportation systems, offering convenient and flexible transportation options for passengers.
} 
By leveraging digital technology and efficient dispatching methods, such as app-based booking and real-time driver allocation, these services have changed the way individuals travel within cities \cite{tirachini2020ride}. 

As these services continue to evolve, several new research questions from different fields arise. For fleet operators and platform providers, one important objective of operating a successful ride-hailing service is to keep the cruising time and distances without passengers as low as possible. The decisions made before and after each ride, such as driver repositioning, can significantly impact the overall efficiency and profitability of the service. In this context, dispatching, which refers to the process of matching passenger requests with available drivers, can intelligently pair passengers and drivers based on factors like proximity of pick-up locations, destination alignment and potential upcoming ride-sharing opportunities after the ride. Additional impact is expected by rebalancing strategies, which decide on where to send vehicles when they are not in use, e.g., after they have completed a ride.

However, ride-hailing services can be limited in their ability to implement post-trip strategies similar to those utilized by traditional taxis due to regulatory requirements. 
For example, in Germany, ride-hailing services are currently obligated to send their vehicles back to their point-of-business after a ride is completed and no follow-up order has been issued.
Such regulations do not only affect the efficiency and profitability of ride-hailing services, but also may have negative impacts on general traffic and the environment.

Since Simulation tools have demonstrated their potential to address questions pertaining to traffic and environmental aspects, particularly in conjunction with the utilization of real data, for the work in this paper, we rely on simulation to address the stated questions. Accordingly, in this paper, we introduce a simulation-based methodology that leverages the Eclipse MOSAIC simulation framework and the extensive city traffic scenario BeST to assess rebalancing strategies for ride-hailing services.
As the first step, we discuss the aspects, which need to be regarded for simulation modelling when analysing such a system. 
Second, we present an exemplary study analyzing and comparing different rebalancing strategies.
The range of strategies that could be used here is huge and their impact can be greatly influenced by the individual business of each company. 
With our simulation-based methodology, we can show how even basic post-ride strategies can have great potential for improvement compared to currently implemented public policy-based mechanisms.

In this paper, we provide an overview of related work (Section~\ref{sec:related_work}), followed by a detailed explanation of aspects needed to integrate real data from existing ride-hailing services for simulation modelling using Eclipse MOSAIC (Section~\ref{sec:methodology}). Section~\ref{sec:experiments} and Section~\ref{sec:results} are dedicated to present an exemplary study, investigating different rebalancing strategies. Section~\ref{sec:discussion} gives a tangible interpretation of the results. Finally, conclusions are drawn in Section~\ref{sec:conclusion}.


\section{Related Work}
\label{sec:related_work}

The majority of research on ride-hailing and ride-pooling concentrates on optimizing dispatching and rebalancing algorithms towards an economical efficient operation.
Most commonly, the goal is to reduce waiting times of customers, or to minimize the empty cruising time, that is, reducing the empty pick-up mileage by integrating intelligent dispatching, and/or reducing the empty rebalancing mileage by sending vehicles to specific locations after a ride has been finished.
Many of these ride-hailing algorithms are based on linear programming, which often does not take into account the dynamic nature of on-demand ride-hailing, where passengers and drivers can hop on and off at any time \cite{yuan2024}. 
Instead, current dispatching algorithms follow a more greedy approach, in which the current situation is considered and optimized rather than the entire problem on a global scale \cite{makhdomi2024greedy}. 
Utilizing reinforcement learning can further improve such methods \cite{yuan2024}.

Besides assigning vehicles to customers (dispatching), the relocation of empty vehicles in the ride-hailing system is of crucial importance for an efficient operation. 
These rebalancing algorithms mostly follow a demand driven approach, by moving idle vehicles to locations with potentially higher demand.
Next to novel approaches utilizing reinforcement learning \cite{deng2022}, there are further approaches which not only focus on operational efficiency, but also on fairness for customers \cite{guo2023fairnessenhancing, yuan2024}.
Many approaches furthermore combine ride-matching and rebalancing in an integrated approach \cite{guo2021, tuncel2023}.
Nevertheless, in some cases public policies may limit the implementation of rebalancing strategies.
For instance, in Germany ride-hailing vehicles are not allowed to cruise after a ride or strategically target interesting points for future passengers (i.e., locations of high demand).
Instead, they are obligated to return to the point-of-business and wait there until a new order is issued (§ 46 Abs. 2 Nr. 1 PBeFG, \cite{pbefg}).

A suitable approach to study the efficiency or impacts of ride-hailing algorithms is the application of simulation.
The main advantage of the simulation approach is that not only the dynamic nature of the ride-hailing system itself, but also additional aspects of the system such as background traffic and communication aspects can be considered.
In a further step, holistic simulations could also be used as a digital twin of the real system, in order to test ITS solutions (such as ride-hailing systems) intensively before applying them to the field (software-in-the-loop) \cite{schrab2022tits}.

Simulation-based studies on ride-hailing and ride-pooling have already been done using open-source tools such as MATSim \cite{matsim2016}.
These studies mostly concentrate on the operation and efficiency of ride-hailing and ride-pooling systems \cite{zwick2020}, or the feasibility of a city-wide deployment of autonomous robo-taxis \cite{hoerl2017}.
MATSim utilizes queue-based models (which makes the simulations very fast), 
however, the discretized space resolution of MATSim limits the modeling of ITS solutions based on perception (e.g., \cite{protzmann2022}), which we plan to investigate further using the presented ride-hailing model.

On the other side of open-source software is the microscopic traffic simulation tool Eclipse SUMO which uses car-following models in the continuous space \cite{sumo2018}. 
SUMO already comes with the possibility to integrate emission models based on HBEFA \cite{emissions2015, hbefa}.
Furthermore, SUMO supports demand responsive transport (DRT) by implementing taxi devices, which are able to model ride-hailing systems to some extent.
However, for the presented work this approach was not customizable enough and rebalancing strategies could not be added without diving into the core of SUMO.

Summarizing, most of the research on ride-hailing focuses on optimizing operation.
However, simulation-based investigations of ride-hailing strategies on environmental impacts, such as emissions, especially on a large city-wide scale, are unknown to us.
With this paper, we aim to fill this gap by implementing a ride-hailing model for the open-source co-simulation framework Eclipse MOSAIC \cite{schrab2022tits}, which already includes coupling to SUMO. Thanks to its modular architecture and versatile API it is very easy to implement custom models and applications which interact with SUMO by adding low computational overhead \cite{protzmann2022}.
\section{Methodology}
\label{sec:methodology}

In this section we present our methodology for analysing ride-hailing services utilizing a suitable simulation environment.
Usually, in traffic analysis, origin-destination (OD) matrices are used to describe the traffic demand which is then applied in a step called "traffic assignment" by finding a departure time and a vehicle route trough the road network.
This method fits very well when analysing vehicle traffic generated by the population of a city.
For modeling demand responsive transport (DRT) such as ride-hailing, however, this static approach is not sufficient.
The demand (passenger requests) is highly dynamic, since passengers join and leave the system at any time and space.
Furthermore, the supply (vehicle assignment) depends on various aspects, such dispatching or rebalancing decisions, and background traffic which influences travel times and route choice.
It is therefore necessary to include data of individual rides, including times and locations of pick-up and drop-off events.
For our analysis we utilize such ride data provided by a middle-sized ride-hailing fleet operator from Berlin, Germany.
This data is similar to the popular TLC and FHV dataset \cite{taxiNyc}, but contains accurate position data of pick-up and drop-off locations instead of zones.
We use this dataset to feed a ride-hailing model with dynamic ride orders, and we apply this model on top of a large-scale traffic scenario which covers 24 hours of motorized individual transport in Berlin.

\subsection{Input data analysis}

In order to evaluate ride-hailing services realistically, data about fleet behavior is essential.
This includes the individual trips of the vehicles, given by location and time of departures and arrivals at pick-up and drop-off points.
For this work, we received extensive data in form of logbooks of a middle-sized fleet (around 50 vehicles) for the period January to October 2023.
The provided dataset contains information about over \SI{120000} ride orders within the City of Berlin and its surrounding areas, such as Potsdam and the BER airport. 
For each order, the following fields have been provided:

\begin{itemize}
    \item Time when the ride order has been issued by the customer.
    \item The vehicle which executed the ride.
    \item Location of the vehicle when it accepted the ride order.
    \item Time and location at which the customer was picked up.
    \item Time and location at which the customer was dropped off after the ride finished.
\end{itemize}

Before applying the data and utilizing it as an input for our experiments, we analyze the input data extensively.
For this initial analysis we do not conduct any simulations.
Instead, we apply statistical methods and use fastest path calculations through an empty road network when analyzing mileages.

To understand the data, we study how the ride-hailing service which provided the data actually operates.
Drivers start their shift at the point-of-business (PoB) of the operator, which is located in the city center near Potsdamer Platz in our case.
A driver accepts incoming ride orders and immediately drives to the pick-up location.
During the ride, the driver can already accept a follow-up order.
After finishing the ride and dropping-off the customer, there are two possible scenarios: 
If a follow-up order has already been accepted during the previous ride, the driver will head to the next pick-up location.
If no follow-up order is existing, the driver is obligated to return directly to the PoB.
The driver can accept new ride orders during that trip to PoB and can then change the target to the new pick-up location.
Otherwise, the driver will wait at the PoB until a new ride order is issued.

In Fig.~\ref{fig:follow_up_orders} we inspect the amount of these different scenarios to happen.
In \SI{40}{\percent} of the cases, the drivers accept a follow-up order already during the current ride.
In all other cases, the drivers are obliged to return to the PoB after dropping off the customer of their current ride. 
During this return trip, new ride orders can be accepted, which happens in \SI{50}{\percent} of all cases.
In \SI{10}{\percent} of the cases, the driver accepts a new ride while waiting at the PoB.
This proportion, however, varies over the days of a week.
At working days the proportion of rides with direct follow-up orders is much lower than on weekends.
For example, on a Wednesday in average only \SI{31}{\percent} of rides end with an accepted follow-up ride, while on weekends this ratio increases to up to \SI{48}{\percent}, mostly due to higher workload.
As a result, there are more and longer return trips on working days than on weekends.
The major drop on Tuesday is produced by maintenance and non-operation of the fleet.

\ifthenelse{\boolean{IEEE_ACCESS_FORMAT}}
{
\Figure[h!](topskip=0pt, botskip=0pt, midskip=0pt)[width=0.99\columnwidth]{figures/plots/30_follow_up_orders}
{ \textbf{Total number of rides per day of week and the moment when the follow-up ride got accepted.}\label{fig:follow_up_orders}}
}
{
\begin{figure}[htbp]
    \centering
    \includegraphics[width=0.99\columnwidth]{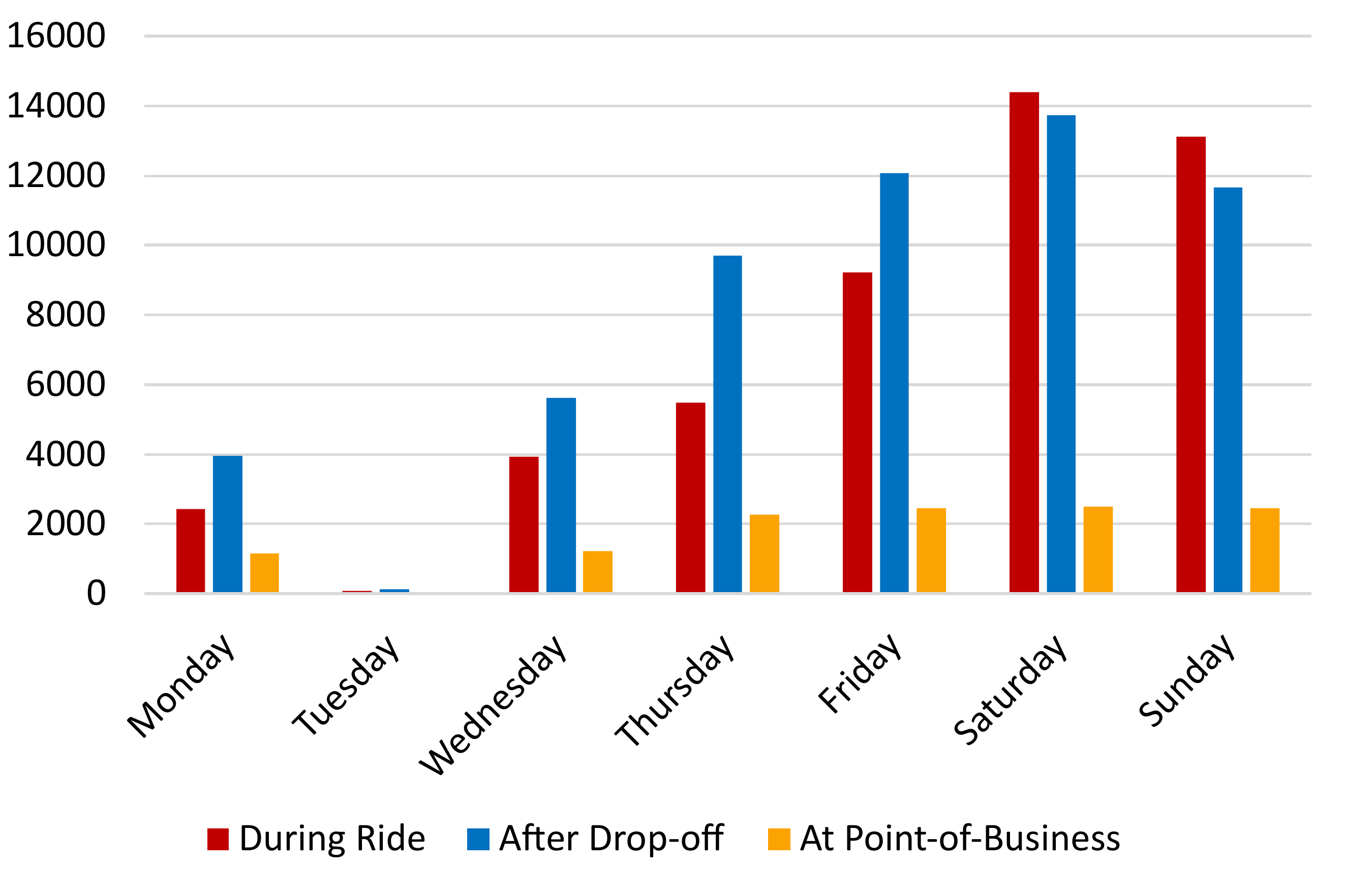}
    \caption{Total number of rides per day of week and the moment when the follow-up ride got accepted.}%
    \label{fig:follow_up_orders}
\end{figure}
}

\ifthenelse{\boolean{IEEE_ACCESS_FORMAT}}
{
\Figure[h!](topskip=0pt, botskip=0pt, midskip=0pt)[width=0.99\columnwidth]{figures/plots/30_methodology_travel_distances}
{ \textbf{Proportion of the average mileages the vehicles travel at each day categorized by their reason.}\label{fig:methodology_travel_distances}}
}
{
\begin{figure}[htbp]
    \centering
    \includegraphics[width=0.99\columnwidth]{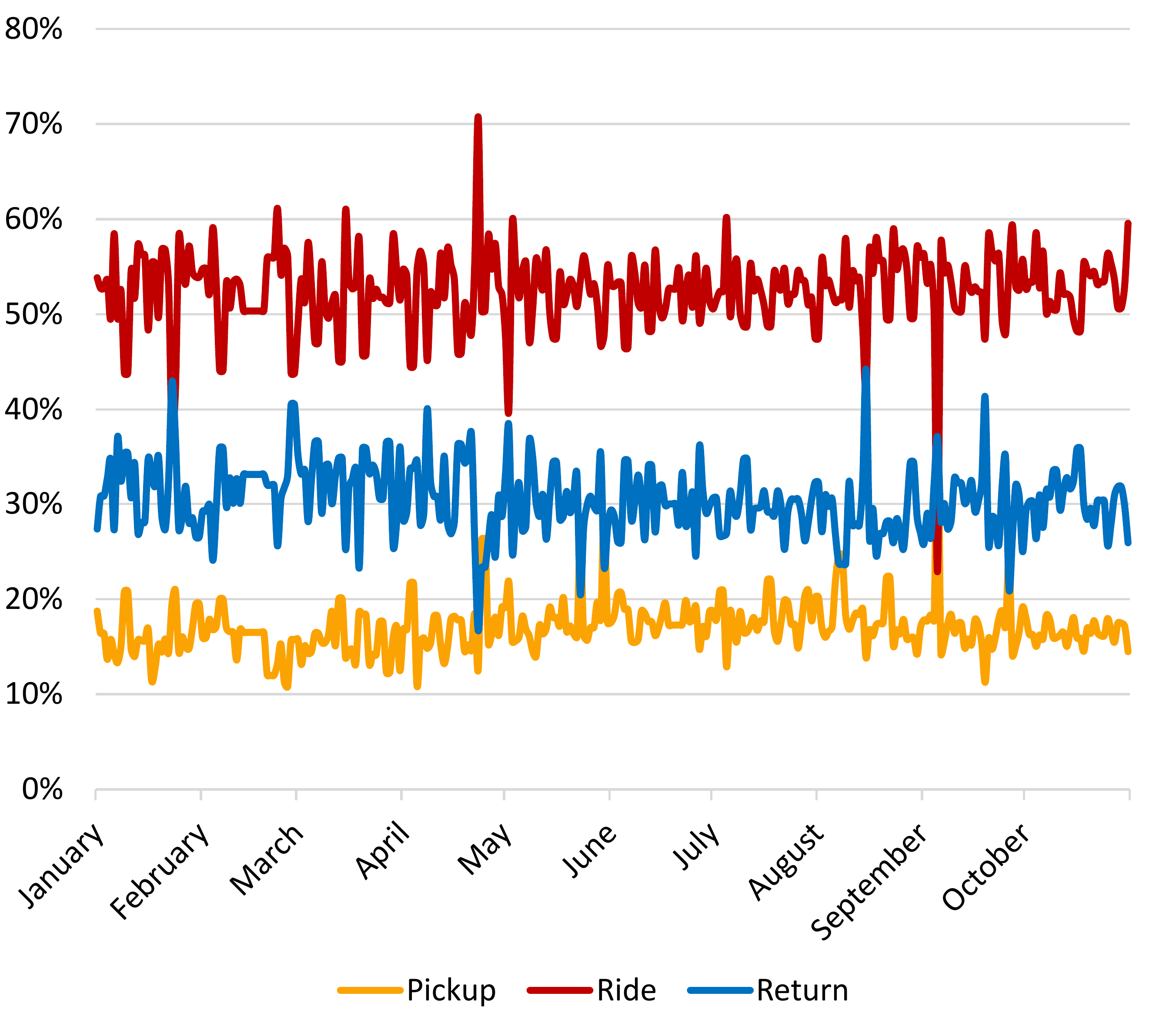}
    \caption{Proportion of the average mileages the vehicles travel at each day categorized by their reason.}%
    \label{fig:methodology_travel_distances}
\end{figure}
}
Ride operation can be divided in three different categories, when looking at the reason a driver is moving around \cite{hoerl2017fleet}.
The \textbf{pick-up mileage} is produced, when the driver is on the way to the customers pick-up location. 
The \textbf{customer mileage} or \textbf{ride mileage} is produced during the actual ride, bringing the customer from pick-up to drop-off location. 
After that, if no follow-up order is already dispatched, the \textbf{rebalancing mileage} is produced by driving to a suitable position until a new ride order is accepted.
In our particular case, the rebalancing mileage is always produced by returning back to the PoB, thus naming it \textbf{return mileage} in the following.
In Fig.~\ref{fig:methodology_travel_distances} we look at the proportion of the different mileage types the vehicles travel at each day.
It can be clearly seen, that on average only \SI{52}{\percent} of the daily mileage of the fleet is serving for the actual rides.
Drivers furthermore spend \SI{17}{\percent} of their driving for the way to the pick-up locations of the customers.
The return mileage in our example shows a rather high proportion of about \SI{31}{\percent} in average.

In this initial analysis of the input data two important key messages can be found.
Firstly, the order situation varies a lot over weekdays, affecting directly the driving behavior after customers have been dropped-off at their destination.
Secondly, return mileage plays a major role in the given fleet data, which opens room for strategies on how these can be reduced.
Just for this fleet, the total accumulated distance of return trips is estimated from the input data with over \SI{250000}{\km}. 
Having in mind that this fleet consists of only \SI{1}{\percent} of the total number of ride-hailing vehicles in Berlin, a huge potential of saving traffic volume and emissions can be assumed.
However, before making detailed and reliable statements, holistic simulations of ride-hailing services and urban traffic is essential and is presented in the next steps.

\subsection{Ride-Hailing Model}

We employ the application simulator of the Eclipse MOSAIC co-simulation framework \cite{schrab2022tits} to implement our ride-hailing model. 
The MOSAIC framework takes the role of coupling and synchronizing the traffic simulator SUMO the with rest of the model. 
Communication between entities (vehicles and ride-hailing service) is modeled using the MOSAIC Cell simulator.
The overall setup is depicted in Fig.~\ref{fig:methodology_mosaic_setup}.
The model itself consist of applications which implement the MOSAIC Application API, realizing the following tasks divided to different applications. An application running on a central server entity takes care of managing ride orders and the assignment to vehicles. Two further applications are mapped to each ride-hailing vehicle, which take care of internal ride and state management, and path finding and stop planning. In detail, the model implements the following tasks:

\begin{itemize}
    \item Generating a logbook with ride-orders from the given input data and queuing ride orders according to the current timestamp of the simulation.
    \item Exchanging information between vehicles and the ride-hailing service via a cellular network for location updates and ride-order assignment.
    \item Assigning ride-orders to individual vehicles. In this work we always take the original assignment from the logbook as we did not implement a dispatching algorithm, but the model is designed to easily implement dispatching algorithms if required.
    \item Calculating routes from vehicle's current to their next stop position (pick-up, drop-off, or point-of-business). Here we consider turn costs to avoid time-consuming left-turns \cite{huebner2015}. 
    \item Initiating halts at the stop position and wait until the passenger enters (pick-up), or exits the vehicle (drop-off), or until a new ride has been issued. The wait time at pick-up and drop-off location has been configured with a minimum time of \SI{30}{\second}.
    \item Applying the chosen rebalancing strategy (see Section~\ref{sec:experiments}), which is executed after the ride has finished and no follow-up order has been issued. As the default strategy, vehicles return to the PoB.
    \item Logging of detailed information about the rides, such as driven distance and travel time per individual trip, waiting times, and several more.
\end{itemize}

\ifthenelse{\boolean{IEEE_ACCESS_FORMAT}}
{
\Figure[t!](topskip=0pt, botskip=0pt, midskip=0pt)[width=0.95\columnwidth]{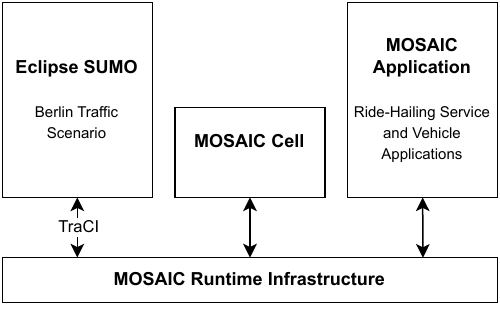}
{ \textbf{The MOSAIC simulation setup includes traffic simulation, the ride-hailing service as an application model, and the cellular simulator to enable communication between vehicles and the ride-hailing service.}\label{fig:methodology_mosaic_setup}}
}
{
\begin{figure}[t!]
    \centering
    \includegraphics[width=0.95\columnwidth]{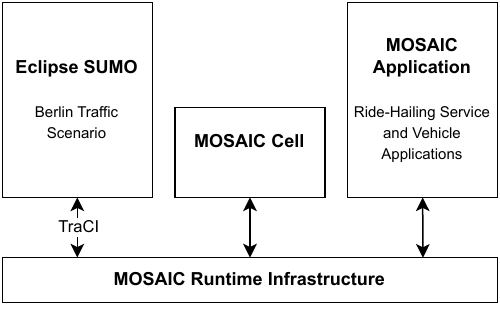}
    \caption{The MOSAIC simulation setup includes traffic simulation, the ride-hailing service as an application model, and the cellular simulator to enable communication between vehicles and the ride-hailing service.}%
    \label{fig:methodology_mosaic_setup}
\end{figure}
}

\subsection{Traffic Scenario}

The presented ride-hailing model works already on its own using a suitable road network. 
However, for a realistic simulation, we integrate general traffic produced by the individuals within the city.
For such simulation, we require a traffic scenario which includes routes and departure times of individual vehicles for a long time frame within the area under research.
For this purpose, we employ the Berlin SUMO Traffic (BeST) scenario \cite{schrab2023best} in our analysis.
This large-scale model developed for SUMO covers 24 hours of motorized individual traffic for the complete city of Berlin.
It includes \SI{2,25} million individual trips which were calibrated against real induction loop data~\cite{schrab2023best}.

A drawback of this scenario is, that it does not contain traffic or road information outside of Berlin.
However, from the input data, ride orders often leave the city to pick-up or drop-off passengers in suburbs, or larger hotspots like the BER airport which is located \SI{5}{\km} outside of Berlin.
As a consequence, rides which end or start outside the modeled area are ignored from the dataset. 
This affects \SI{5}{\percent} of all rides, however, \SI{2}{\percent} of all rides serving BER airport (as depicted in Fig.~\ref{fig:berlin_map}).
Therefore, we decided to extend the BeST scenario by adding the airport, including road data of the BER terminal access, all associated highways, and the connection with the existing roads of the BeST scenario.
Still, removing a single ride from a list of connected orders would disturb the model, as too much time between two subsequent orders would be introduced.
To overcome this, we group subsequent rides which are assigned to the same vehicle with no pause longer than two hours to \textit{shifts}.
If at least one ride of such shift is outside the area of Berlin, we dismiss the whole shift.
Eventually, a realistic model of the general traffic has been created which we use as the base to model our ride-hailing service on top.
In a next step, we validate our model against the provided input dataset.

\ifthenelse{\boolean{IEEE_ACCESS_FORMAT}}
{
\Figure[tb](topskip=0pt, botskip=0pt, midskip=0pt)[width=0.99\columnwidth]{figures/30_berlin_map}
{ \textbf{Map of the model area.} Each dot represents a pick-up or drop-off location from the provided input dataset. 5~\% of all rides start or end outside Berlin, while 2~\% of all rides start or end at BER airport (blue area in the south).\label{fig:berlin_map}}
}
{
\begin{figure}[tb]
    \centering
    \includegraphics[width=0.99\columnwidth]{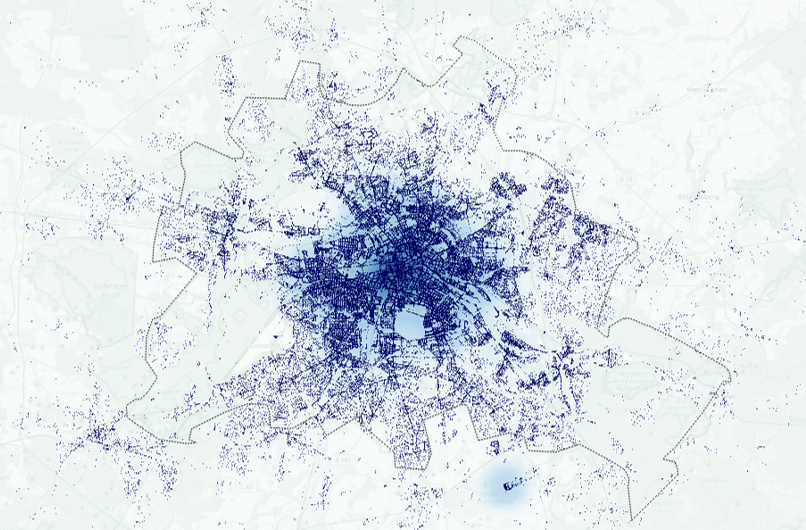}
    \caption{Map of the model area. Each dot represents a pick-up or drop-off location from the provided input dataset. 5~\% of all rides start or end outside Berlin, while 2~\% of all rides start or end at BER airport (blue area in the south}%
    \label{fig:berlin_map}
\end{figure}
}

\subsection{Model Validation}

To validate our model we run initial simulations and compared the results with the given input data.
We generated an artificial logbook based on the given input data (more details on this logbook generation can be found in Section~\ref{sec:experiments}).
For each ride order various properties are written out by the simulation, such as the travel time for the actual ride, the stopping locations, the times at which customers are picked-up or dropped-off, travel distances, and many more.
Some of these measurements can be directly compared with the original logbooks from the input data.
Firstly, we compare the travel times of each ride produced by the simulation in Fig.~\ref{fig:validation_ride_travel_time}.
Most of the rides require similar travel time as given by the real logbook.
\SI{72}{\percent} of the rides show an absolute difference in travel time of under \SI{200}{\second}, while the median can be found at \SI{-69}{\second} travel time difference. 
Ride-sharing vehicles in the simulation are therefore slightly faster than their real counterparts, however, it must be considered that the routes the vehicles took in reality are not given.
In the simulation, however, vehicles take the fastest route in that moment, which could explain faster rides.
Comparing the times at which customers are picked-up, gives a slightly broader distribution (see Fig.~\ref{fig:validation_pickup}). 
While the median can be found at only \SI{3}{\second} difference, still \SI{58}{\percent} of the pick-up times are below \SI{200}{\second}.

However, the provided dataset does not give the complete picture of what happens during a ride.
For example, some rides showed a discrepancy between ride length and travel time, such that they show a rather long time between pick-up and drop-off, but also a rather short distance between both of the locations.
Furthermore, it is not clear if the route choice of the drivers is equal to the route calculation from the simulation.
Necessary tasks, such as taking breaks or refueling at gas stations, are also not part of the data, which all increase the deviations we see in the validation of the model. Nevertheless, the shown (and other) metrics lead us to the conclusion that our model depicts the real data quite well and is therefore sufficient for further, more detailed experiments. 

\ifthenelse{\boolean{IEEE_ACCESS_FORMAT}}
{
\Figure[tb](topskip=0pt, botskip=0pt, midskip=0pt)[width=0.99\columnwidth]{figures/plots/30_validation_ride_duration}
{ \textbf{Distribution of differences in ride travel times produced by the model compared to the input data.}\label{fig:validation_ride_travel_time}}
}
{
\begin{figure}[tb]
    \centering
    \includegraphics[width=0.99\columnwidth]{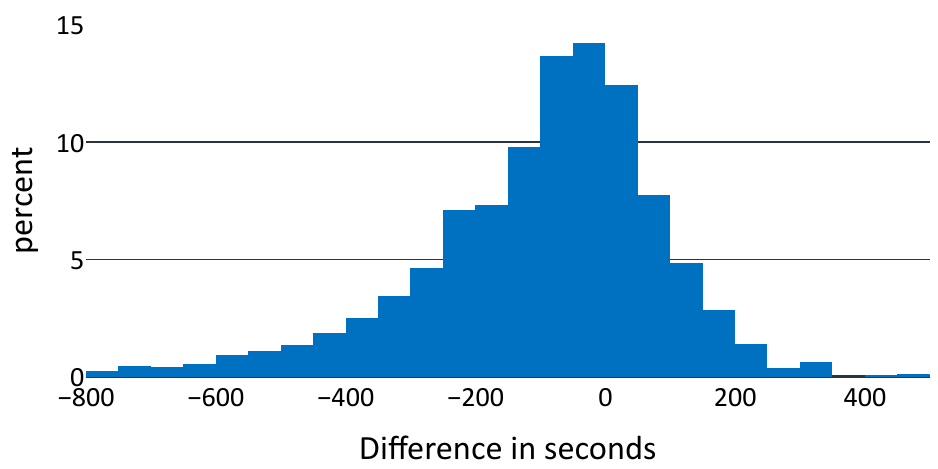}
    \caption{Distribution of differences in ride travel times produced by the model compared to the input data.}%
    \label{fig:validation_ride_travel_time}
\end{figure}
}

\ifthenelse{\boolean{IEEE_ACCESS_FORMAT}}
{
\Figure[tb](topskip=0pt, botskip=0pt, midskip=0pt)[width=0.99\columnwidth]{figures/plots/30_validation_departure}
{ \textbf{Distribution of differences in pick-up time produced by the model compared to the input data.}\label{fig:validation_pickup}}
}
{
\begin{figure}[tb]
    \centering
    \includegraphics[width=0.99\columnwidth]{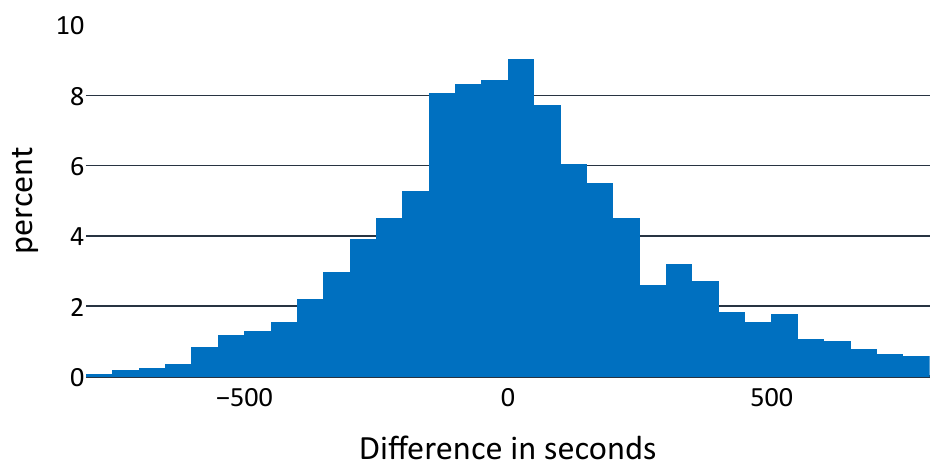}
    \caption{Distribution of differences in pick-up time produced by the model compared to the input data.}%
    \label{fig:validation_pickup}
\end{figure}
}

\section{Experiments}
\label{sec:experiments}

As an exemplary study based on the developed ride-hailing model, we investigate traffic and environmental impacts of deploying different return strategies.
Before conducting any experiments, we explain two key aspects which we use to parameterize the simulations.
Firstly, we introduce a generation of artificial logbooks based on the provided input data.
This allows us to overcome the problem of invalid rides due to outside locations, as explained in Section~\ref{sec:methodology}.
Furthermore, it gives us the possibility to use logbooks for an arbitrary fleet size, and to generate several different logbooks for any day of week.
A second aspect regards the behavior of the drivers after a customer has been dropped-off if no follow-up order has already been issued.
Here we employ different of such rebalancing strategies to measure their impact on the mileage for the ride-hailing operator and emissions of greenhouse gases and air pollutants.

\ifthenelse{\boolean{IEEE_ACCESS_FORMAT}}
{
\Figure[!h](topskip=0pt, botskip=0pt, midskip=0pt)[width=0.95\textwidth]{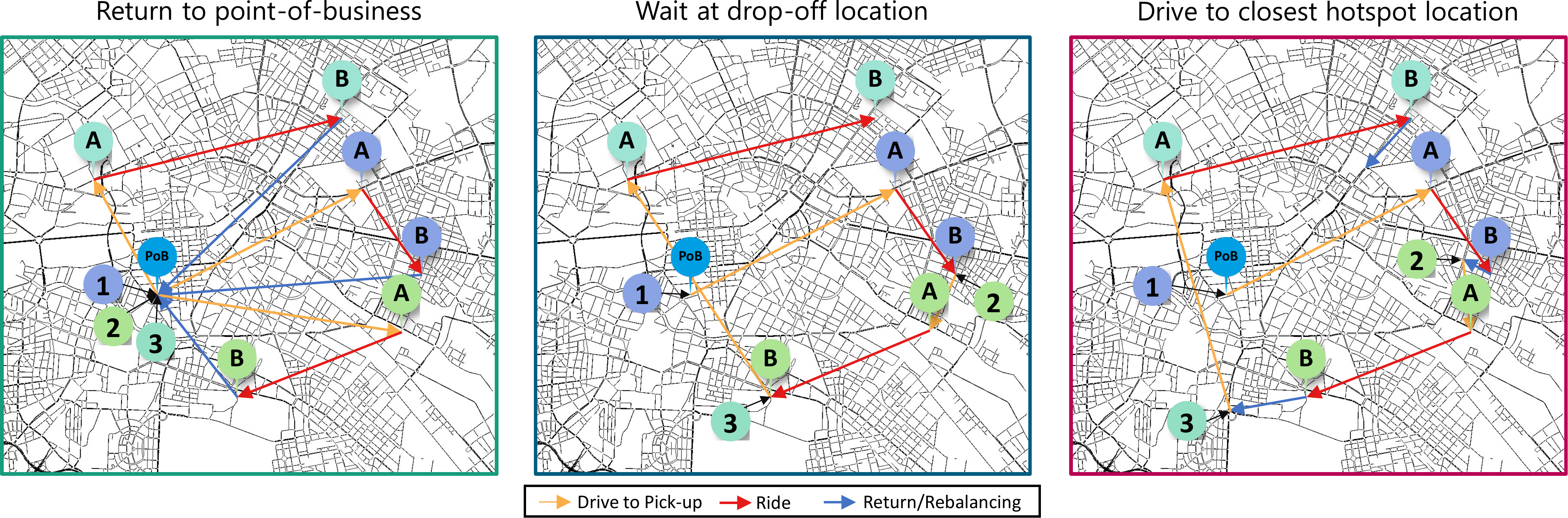}
{\textbf{Three rebalancing strategies implement different behavior for the drivers after they finished a ride without having a direct follow-up order. } The numbers depict the location at which the follow-up ride (same color A->B) was accepted.\label{fig:experiments_return_strategies}}
}
{
\begin{figure*}[tb]
    \centering
    \includegraphics[width=0.95\textwidth]{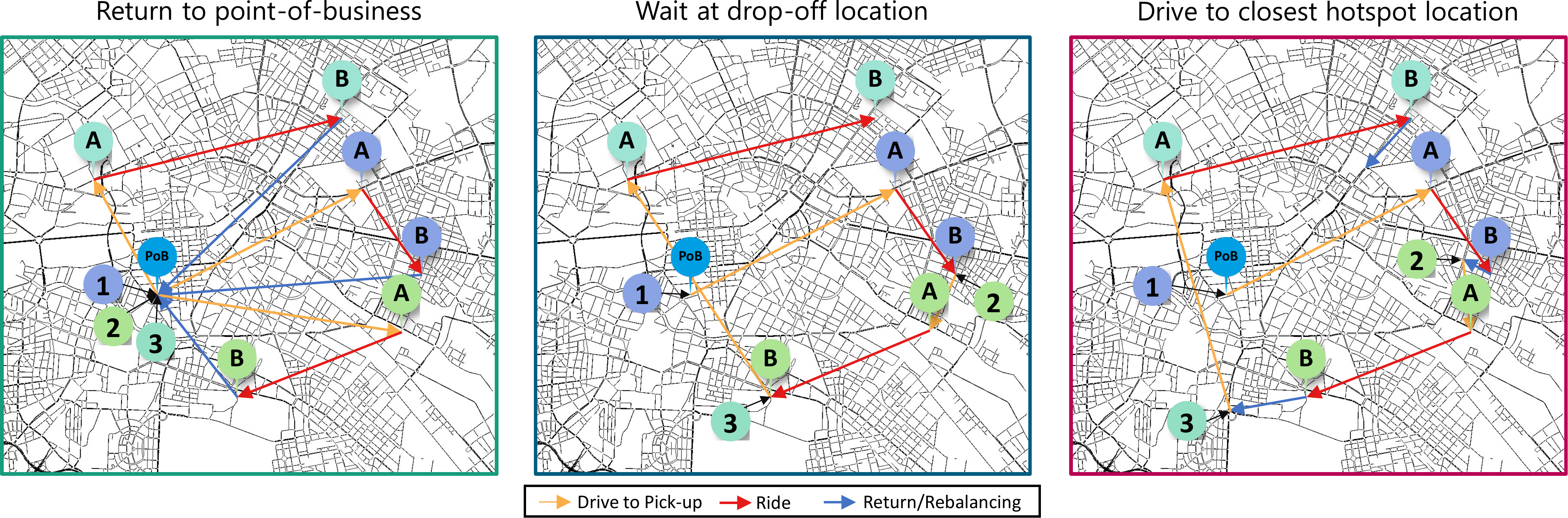}
    \caption{Three rebalancing strategies implement different behavior for the drivers after they finished a ride without having a direct follow-up order. The numbers depict the location at which the follow-up ride (same color A->B) was accepted.}%
    \label{fig:experiments_return_strategies}
\end{figure*}
}

\subsection{Logbook Generation}

The provided input data contains \SI{120000} individual ride orders covering a whole year of operation.
For our experiments, we are interested in modeling typical days, such as a working day or weekend day.
Instead of choosing specific dates, we decided to generate logbooks which model such typical days based on the input data.
In that process, sequences of rides which are related to each other (shifts) are kept together in order to keep the ride-hailing model functioning.
The logbooks are generated by following, given the \textit{day of week} and the required \textit{fleet size}:
\begin{itemize}
    \item Select a random ride from the original logbook which was executed at the given day of week.
    \item Collect all rides which are related to the selected ride (i.e., all rides which have been assigned to the same vehicle within the same shift on that specific day).
    \item Bundle all the selected rides as a shift. This shift is then assigned to the current vehicle, if there is at least a pause of 2 hours to the next shift and the previous shift.
    \item Assign at maximum 3 shifts per vehicle.
    \item Repeat the process until the given \textit{fleet size} is reached.
\end{itemize}

By using different random seeds, this process results in different logbooks, giving us the possibility to generate near endless variants of logbooks which match the typical behavior of the given day of week. 
Furthermore, it is possible to scale or limit the logbook to a larger or smaller fleet size.
However, since road infrastructure may limit the number of vehicles of a fleet to park or wait at the point-of-business, we decided for our experiments not to exceed the initial fleet size of 50 vehicles.
For our experiments we generated logbooks for 2 different days of the week (\textit{Wednesday} and \textit{Saturday}) with 8 random variations each.

\subsection{Rebalancing Strategies}

As an exemplary study of this work, we model different behavior of drivers after they dropped-off the current passengers with the goal to measure impacts on traffic and environment.
In this study, we kept the strategies rather simple, since we focused on the modeling part of the system.
When drivers finish a ride, there are several possibilities, if no follow-up order has been issued:

\begin{itemize}
    \item \textbf{Return to Point-of-Business}: This behavior is currently obligatory for all ride-hailing services operated in Germany. 
    If no follow-up ride has been accepted by the driver, then she has to drive back to the point-of-business, aka headquarters of the operator. 
    It is allowed, though, to drive to the next pick-up point once the next order has been accepted.
    In this work we use this "strategy" as the baseline to find how other strategies can improve operation of the ride-hailing service in terms of saving driving effort and emissions.
    \item \textbf{Waiting}: In this strategy, drivers just wait at their last drop-off location until the next ride order is issued. 
    There is no requirement to drive back to the point-of-business at any time.
    It can be assumed, that this strategy might be not very efficient from an operator point-of-view, as future customers are most likely not expected at the last drop-off location.
    Especially, waiting within residential or sub-urban areas is probably not helpful.
    \item \textbf{Drive to Hotspot}: This strategy is somewhere between the top mentioned and is similar to the relocation zone approach presented \cite{gueriau2020}. 
    For this strategy we chose 60 different locations within Berlin which can be expected to have a higher probability of being close to a pick-up location.
    These so-called "hotspots" are derived from the complete input data, by doing a spatial clustering analysis on all existing pick-up locations, using the \mbox{DBSCAN} algorithm \cite{ester1996}.
    After dropping-off the passenger, the vehicles drive to the closest hotspot to wait there instead, until the follow-up order is issued.
    In only \SI{20}{\percent} of the cases, drivers still wait at their latest drop-off location.    
\end{itemize}

An overview about the different strategies is depicted in Fig.~\ref{fig:experiments_return_strategies}.
It furthermore must be noticed, that the assignment and order of rides is not affected by the presented strategies.
This would have required to implement complex dispatching algorithms, which we see out of scope here.
Furthermore, such dispatching would require to replicate the baseline behavior very closely in order to receive comparable results.
This, on the other hand, would require us most probably to include business secrets into our model.
Instead, we decided to use the very same assignment for each strategy.
In future work, we plan to integrate state-of-the-art dispatching and rebalancing algorithms.

\subsection{Simulation Series}\label{subsec:simulation_series}
For the exemplary study, we run 48~full-day simulations resulting from the crossing of each return strategy with each chosen day of week (Wednesday / Saturday) with each of the 8 variations.
Using the SimulationRunner tool of the extended version of \mbox{MOSAIC}, we were able to comfortably parameterize and execute the simulation runs. 
In total, the execution of the entire simulation series took about 6~days on an Intel\textsuperscript{\tiny\textregistered}
 Xeon\textsuperscript{\tiny\textregistered} CPU E5-2623 v3 @ \SI{3}{\giga\hertz}, running three simulations in parallel.

\section{Results}
\label{sec:results}
Using the output from the simulations described in Section \ref{subsec:simulation_series}, we conducted an extensive evaluation on the traffic and environmental impacts of the defined rebalancing strategies.
For the defined rebalancing strategies, we are using the name placeholders \emph{Return} (Baseline), \emph{Wait}, and \emph{Hotspot} throughout this section.

The results for the most relevant KPIs (Key Performance Indicators) from the simulation series are displayed Table~\ref{tab:results}.
These results are averaged over all simulation runs for the respective strategies.
One has to keep in mind that the presented results are comprised of a fleet of 50~vehicles,  

\begin{table*}[t!]
\caption{\textbf{Averaged KPI results from simulation study. Relative improvements compared to Baseline scenario are captured inside the brackets.}}
\label{tab:results}
\centering
\setlength{\tabcolsep}{3.5pt}
\begin{tabular}{|L{0.175\textwidth}|C{0.01\textwidth}R{0.047\textwidth}C{0.01\textwidth}R{0.054\textwidth}L{0.066\textwidth}R{0.054\textwidth}L{0.066\textwidth}|C{0.01\textwidth}R{0.047\textwidth}C{0.01\textwidth}R{0.054\textwidth}L{0.066\textwidth}R{0.054\textwidth}L{0.066\textwidth}|}
\hline
\multicolumn{1}{|c|}{} & \multicolumn{7}{c|}{\textbf{Wednesday}} & \multicolumn{7}{c|}{\textbf{Saturday}} \\
\multicolumn{1}{|c|}{} & \multicolumn{3}{c}{Return} & \multicolumn{2}{c}{Wait} & \multicolumn{2}{c|}{Hotspot} & \multicolumn{3}{c}{Return} & \multicolumn{2}{c}{Wait} & \multicolumn{2}{c|}{Hotspot}  \\
\hline

Total Mileage [\si{\kilo\meter}] & & \num{14724} & & \num{11127} & (\SI{-24}{\percent}) & \num{12060} & (\SI{-18}{\percent}) & & \num{23037} & & \num{18354} & (\SI{-20}{\percent}) & \num{19739} & (\SI{-14}{\percent}) \\

Mileage \emph{Rebalancing} [\si{\kilo\meter}] & & \num{4063} & & \num{0} & (\SI{-100}{\percent}) & \num{889} & (\SI{-78}{\percent}) & & \num{5266} & & \num{0} & (\SI{-100}{\percent}) & \num{1303} & (\SI{-75}{\percent}) \\

Mileage \emph{Pickup} [\si{\kilo\meter}] & & \num{3340} & & \num{3812} & (\SI{+14}{\percent}) & 3855 & (\SI{+15}{\percent}) & & \num{5569} & & \num{6155} & (\SI{+11}{\percent}) & \num{6235} & (\SI{+12}{\percent}) \\

Mileage \emph{Ride} [\si{\kilo\meter}] & & \num{7321} & & \num{7315} & ($\sim$\,\SI{0}{\percent}) & \num{7316} & ($\sim$\,\SI{0}{\percent}) & & \num{12203} & & \num{12200} & ($\sim$\,\SI{0}{\percent}) & \num{12202} & ($\sim$\,\SI{0}{\percent}) \\
\hline

Total $\text{CO}_\text{2}$ Emission [\si{\kilo\gram}] & & \num{1519} & & \num{1137} & (\SI{-25}{\percent}) & \num{1245} & (\SI{-18}{\percent}) & & \num{2382} & & \num{1883} & (\SI{-21}{\percent}) & \num{2038} & (\SI{-14}{\percent}) \\ 

$\text{CO}_\text{2}$ Emission [\si[per-mode=symbol]{\gram\per\kilo\meter}] & & {\num{103.16}} & & \num{102.18} & ($\sim$\,\SI{0}{\percent}) & \num{103.23} & ($\sim$\,\SI{0}{\percent}) & & \num{103.40} & & \num{102.60} & ($\sim$\,\SI{0}{\percent}) & \num{103.25} & ($\sim$\,\SI{0}{\percent}) \\ 

Total $\text{CO}$ Emission [\si{\gram}] & & \num{5912} & & \num{2822} & (\SI{-25}{\percent}) & \num{3098} & (\SI{-18}{\percent}) & & \num{4678} & & \num{5069} & (\SI{-21}{\percent}) & \num{3766} & (\SI{-14}{\percent}) \\

Total $\text{NO}_\text{x}$ Emission [\si{\gram}] & & \num{25.54} & & \num{19.01} & (\SI{-26}{\percent}) & \num{20.45} & (\SI{-20}{\percent}) & & \num{38.58} & & \num{30.35} & (\SI{-21}{\percent}) & \num{32.43} & (\SI{-16}{\percent}) \\

Total $\text{PM}_\text{x}$ Emission [\si{\gram}] & & \num{80.99} & & \num{61.11} & (\SI{-25}{\percent}) & \num{66.14} & (\SI{-18}{\percent}) & & \num{126.20} & & \num{100.39} & (\SI{-20}{\percent}) & \num{107.87} & (\SI{-15}{\percent}) \\
\hline

Mileage per Shift [\si{\kilo\meter}] & & \num{139} & & \num{105} & (\SI{-24}{\percent}) & \num{114} & (\SI{-18}{\percent}) & & \num{164} & & \num{130} & (\SI{-20}{\percent}) & \num{140} & (\SI{-14}{\percent}) \\

Shift Duration [\si{\hour}] & & \formattime{4}{1}{0} & & \formattime{3}{2}{0} & (\SI{-24}{\percent}) & \formattime{3}{20}{0} & (\SI{-17}{\percent}) & & \formattime{4}{47}{0} & & \formattime{3}{49}{0} & (\SI{-20}{\percent}) & \formattime{4}{8}{0} & (\SI{-14}{\percent}) \\
\hline
\end{tabular}
\end{table*}

\subsection{Traffic Impacts}\label{subsec:traffic_impacts}
In the traffic domain, the most important KPIs are typically mileages and travel times.
In Fig.~\ref{fig:results_fig_dist_per_shift} we plot the averaged mileage per shift, both for Saturday and Wednesday.
Additionally, we added box plots to showcase the large variability in shift lengths, which is simply due to the fluctuating nature of ride hailing services.
Comparing the results of the two baseline scenarios, it can be seen that on average mileages per shift are about \SI{25}{\kilo\meter} (\SI{15}{\percent}) longer on Saturday than on Wednesday.
This is likely explained by higher demand on weekends as more people rely on ride hailing services getting to and leaving leisure activities.
Furthermore, drivers are inclined to do longer shifts on weekends as the potential profit is higher.

When comparing the \emph{Wait} and \emph{Hotspot} strategies with the baseline scenario, it is apparent that both on Wednesday and Saturday an reduction of total mileage is possible, with the \emph{Wait} strategy outperforming the \emph{Hotspot} strategy in this regard.
However, this comes with little surprise as we replace the potentially long routes back to the point-of-business with shorter or even no routes.
Interestingly though, the potential gain is larger on Wednesday compared to Saturday.
On Saturday mileage can be reduced by about \SI{20}{\percent} using the \emph{Wait} strategy and \SI{14}{\percent} using the \emph{Hotspot} strategy, while on Wednesday this reduction increases to \SI{24}{\percent} and \SI{18}{\percent}, respectively.
Again, this is likely to the different demands on weekdays and weekends.
Drivers are more likely to receive new bookings faster when the demand is high, leading to shorter return ways.

The previous observations can lead to the conclusion that total mileages for ride hailing fleets can be reduced by \SIrange{20}{24}{\percent}, or \SIrange{14}{18}{\percent} for a more sensible rebalancing strategy.
However, this is not the entire picture, as assigned bookings from the provided log books to the same vehicles regardless of the applied strategy.
Due to this simplification we observed longer average mileages to the next pickup point, for the \emph{Wait} and \emph{Hotspot} strategies.

In an effort to further analyze how the mileages are made up, we grouped all distances by reason (Fig. \ref{fig:results_fig_dist_reason}).
Here we are only considering Wednesday as the results for Saturday show a similar pattern, with increased mileage.
Comparing the average total mileage with the mileage per shift (Fig. \ref{fig:results_fig_dist_per_shift}), it is apparent that the same relative improvements of \SI{24}{\percent} (\emph{Wait}) and \SI{18}{\percent} (\emph{Hotspot}) persist.

Looking more detailed into the distribution of mileages, we can see that the \emph{Ride} mileages are mostly constant ($\sim$\,\SI{7317}{\kilo\meter}) for all strategies, which is not surprising as the bookings for all strategies are the same.
The slight deviation may be explained by differing routes or incomplete rides but will be disregarded.

Of course, the rebalancing mileages show the highest deviations as these are the distances mostly influenced by the applied rebalancing strategies.
With the \emph{Wait} strategy there is no rebalancing mileage as vehicles wait at the last drop-off location until receiving the next booking.
The rebalancing mileage for the \emph{Hotspot} strategy is made up of the distances that vehicles drive to the next synthetic hotspot and are still much smaller than the \emph{Baseline} scenario.
Now due to the aforementioned simplification of the dispatching algorithm, we actually see an increase in the \emph{Pickup} mileage for the novel rebalancing strategies.
Realistically, these mileages should not increase significantly.
In an improved simulation, vehicles would not simply follow the predetermined bookings but rather be dynamically assigned to new bookings depending on their position.
However, with the provided data such dynamic dispatching could not be included within this examination.
With this in mind, the previous potential reductions of \SIrange{20}{24}{\percent} (\emph{Wait}), and \SIrange{14}{18}{\percent} (\emph{Hotspot}) could even be extrapolated to the entirety of rebalancing distances.
Consequently, in the \emph{Baseline} scenario \SI{27.6}{\percent} (\SI{4063}{\kilo\meter}) of all distances are rebalancing distances (i.e., driving to the point-of-business).
Leading to a potential reductions of \SIrange{22.9}{27.6}{\percent} (\emph{Wait}), and \SIrange{17.2}{21.6}{\percent} (\emph{Hotspot}).

\ifthenelse{\boolean{IEEE_ACCESS_FORMAT}}
{
\Figure[t](topskip=0pt, botskip=0pt, midskip=0pt)[width=0.99\columnwidth]{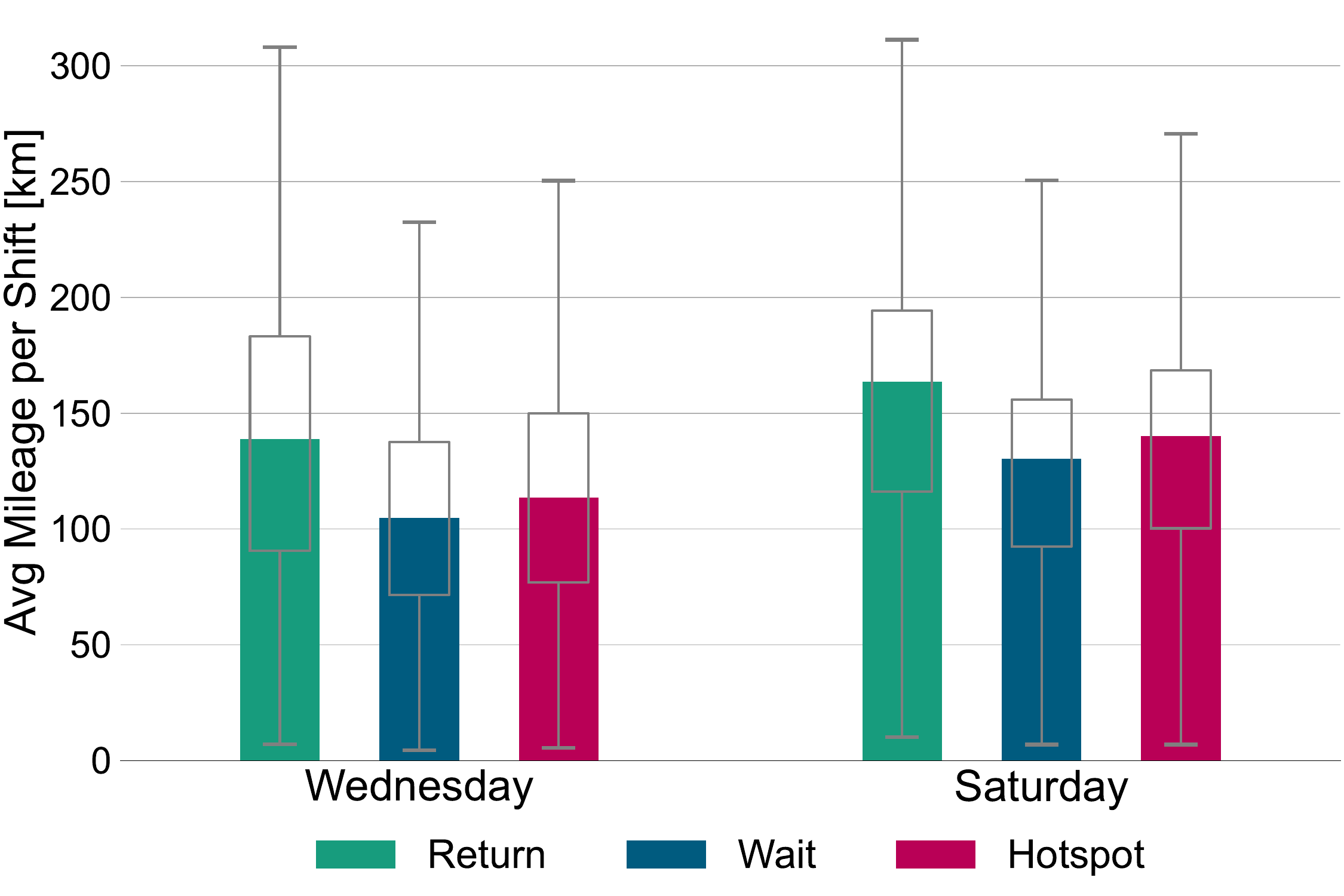}
{\textbf{Comparison of average mileages driven per shift.}\label{fig:results_fig_dist_per_shift}}
}
{
\begin{figure}[t]
    \centering
    \includegraphics[width=0.99\columnwidth]{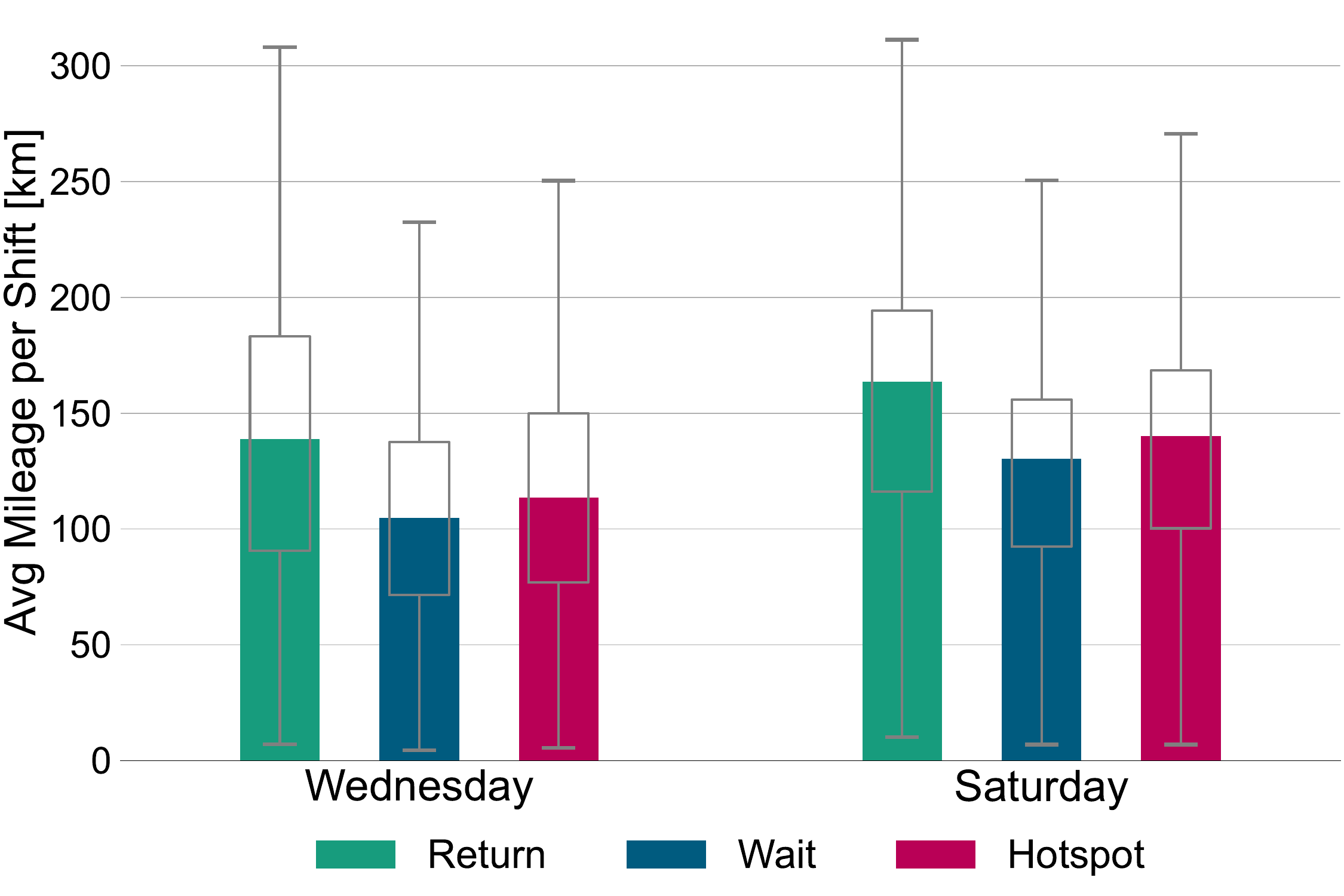}
    \caption{Comparison of average mileages driven per shift.}%
    \label{fig:results_fig_dist_per_shift}
\end{figure}
}

\ifthenelse{\boolean{IEEE_ACCESS_FORMAT}}
{
\Figure[t](topskip=0pt, botskip=0pt, midskip=0pt)[width=0.99\columnwidth]{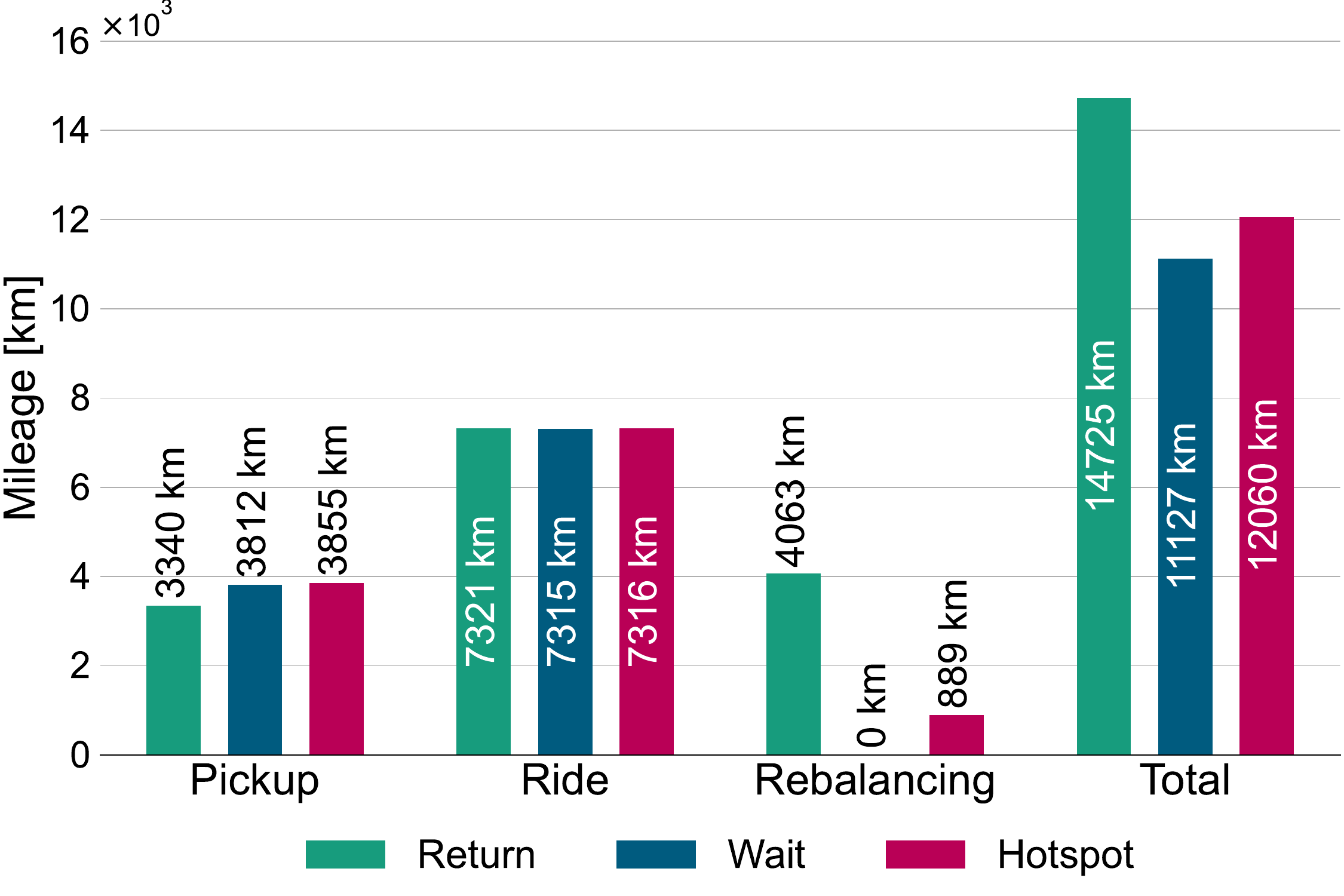}
{\textbf{Wednesday: Split of total distances driven by reason.}\label{fig:results_fig_dist_reason}}
}
{
\begin{figure}[t]
    \centering
    \includegraphics[width=0.99\columnwidth]{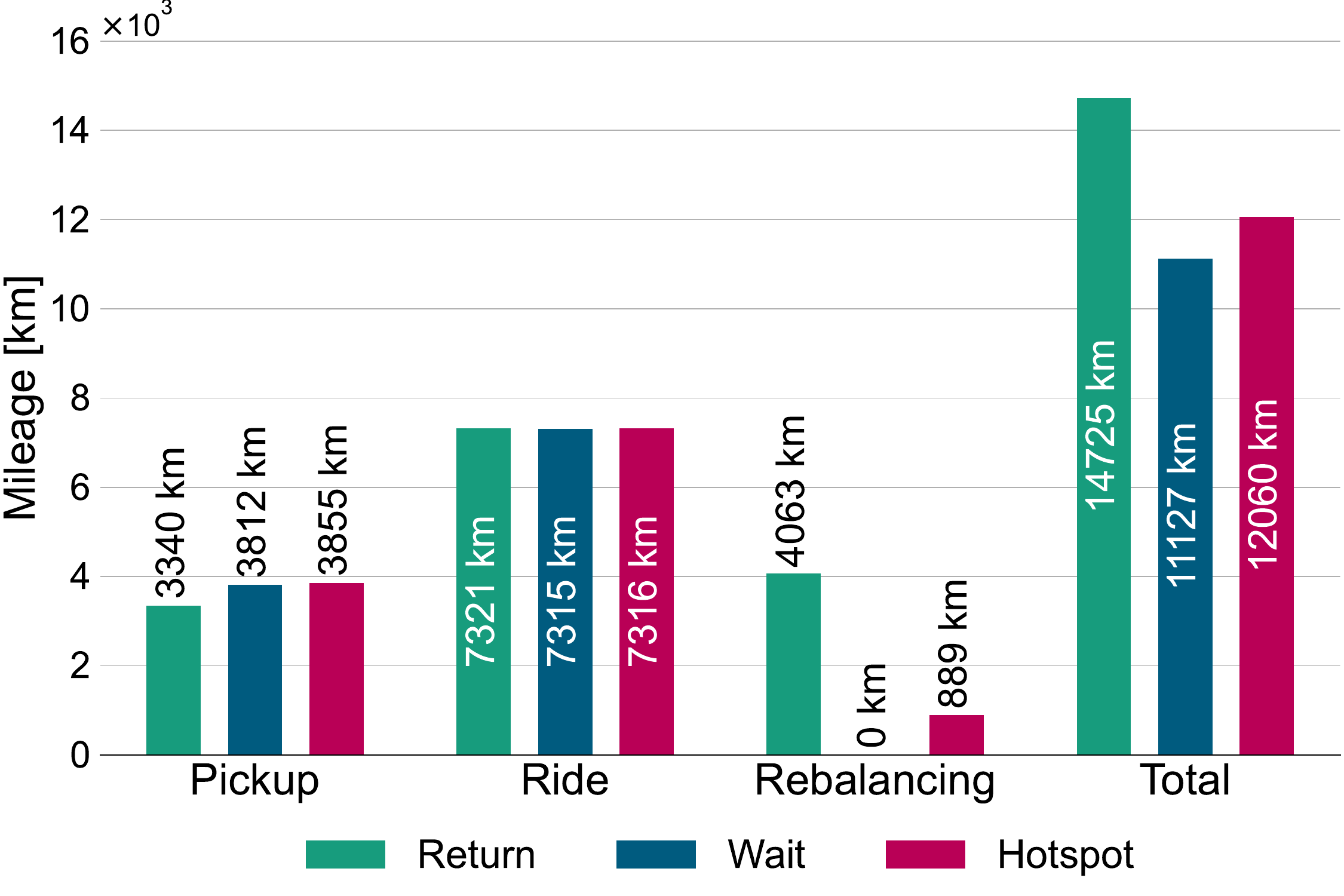}
    \caption{Wednesday: Split of total distances driven by reason.}%
    \label{fig:results_fig_dist_reason}
\end{figure}
}

\subsection{Environmental Impacts}\label{subsec:results_environmental_impacts}
For the evaluation of environmental impacts we built on SUMO's emission model\cite{emissions2015}, which incorporates the latest HBEFA v4.2 \cite{hbefa} for the calculation of fuel consumption and different pollutant emissions.
Specifically, we track Carbon Dioxide ($\text{CO}_{\text{2}}$), Carbon Monoxide ($\text{CO}$), Nitric Oxides ($\text{NO}_{\text{x}}$), and Particulates ($\text{PM}_{\text{x}}$) of varying diameters, mostly \SI{10}{\micro\meter}.
We assumed a uniform fleet of plugin hybrid vehicles with the emission class PHEV, Euro6d (Gasoline), which at the time of writing was the most prominent vehicle type in the provided logbooks.
Vehicles of the Euro6d emission class emit about \SI{100}{\gram} of $\text{CO}_{\text{2}}$ per kilometer.
Averaged emissions for all strategies on both Wednesday and Saturday are shown in Table~\ref{tab:results}.

Initially, these results can be used for a rough re-validation of the applied emission model.
The resulting $\text{CO}_{\text{2}}$ emissions per kilometer for all simulations end up to around \SI[per-mode=symbol]{103}{\gram\per\kilo\meter} (vs. specified \SI[per-mode=symbol]{100}{\gram\per\kilo\meter}), which is well within expected margins. 

Total $\text{CO}_{\text{2}}$ emissions accumulate to \SI{1500}{\kilo\gram} on Wednesday and around \SI{2400}{\kilo\gram} on Saturday.
Applying the proposed rebalancing strategies, we can see that the achievable emission reductions are mostly in line with the reduction in travel distances, with relative deviations of less than \SI{2}{\percent}.
On Wednesday, the \emph{Wait} strategy reduced $\text{CO}_{\text{2}}$ emissions by \SI{382}{\kilo\gram} (\SI{25}{\percent}) and the \emph{Hotspot} strategy by \SI{274}{\kilo\gram} (\SI{18}{\percent}).
In contrast, the absolute savings are larger on Saturday with \SI{499}{\kilo\gram} (\emph{Wait}) and \SI{344}{\kilo\gram} (\emph{Hotspot}).
However, the relative savings are slightly smaller with \SI{21}{\percent}, and \SI{14}{\percent}, respectively.


\section{Discussion}
\label{sec:discussion}
The results from the simulation study indicate that the deployment of different rebalancing strategies can have a large impact on mileages and environmental factors.
In a real-world deployment, these impacts may even increase, as dispatching would improve.
The current implementation makes the simplification of always assigning drivers the same shifts and thereby bookings, regardless of the current position of an vehicle.
Both the \emph{Wait} and \emph{Hotspot} strategy suffer from this, as the \emph{Pickup} mileage increase by \SIrange{11}{15}{\percent} on average, depending on the day and strategy.
In contrast, real-world rebalancing mileage would likely increase compared to the simulation results.
However, consideration of more elaborate dispatching and rebalancing algorithms is out of the scope of this paper.

\subsection{Upscaling Environmental Impacts}

Now, in an effort to quantize the potential savings presented in Section~\ref{sec:results} and give a tangible interpretation of the results,
we tried to extrapolate our findings for the entire ride-hailing traffic for the year 2023 in the city of Berlin.
For this extrapolation some additional prerequisites are necessary as well as some assumptions had to be made.
At the time of writing we had access to logbooks of a 50 vehicle fleet for the months of January 2023 to October 2023.
Thus, two initial scaling factors are required, a) a factor to scale to the entire fleet, and b) a factor scaling results to the months of November and December.
First, for the fleet size we used data from \cite{berlinopendata} which provides monthly updated numbers for taxi and ride-hailing fleet sizes, ranging from \num{4321} to \num{4486}.
Secondly, we assumed that the ride-hailing demand for the last two months would stay similar as the rest of the year and therefore simply scaled our results by a value of $C_y = 1.17$.
Next, we considered the fleet utilization $\rho$ as rarely all vehicles of the fleet are utilized at the same time.
We again extrapolated data from the provided logbooks, where we saw utilization values ranging from around \SIrange{20}{90}{\percent} depending on the day.
Finally, we utilized results from the simulation study, yielding average $\text{CO}_{\text{2}}$ emissions of \SI[per-mode=fraction]{102.35}{\gram\per\kilo\meter} for the entire vehicle fleet.

Now, for each of the considered days we extracted average pickup mileages $s_p$, customer mileages $s_r$, and rebalancing mileages $s_b$.
Next, to get an estimate for the fleet size each day, we multiplied the amount of deployed vehicles $n$ within the corresponding month with the fleet utilization $\rho$ for each day.
Finally, we summed results for each of the logged day and applied the scaling factor $C_y$ to end up with total mileage estimates $S_i$ for the entire year of 2023 (see Equation \ref{eq:total_mileages}).

\begin{equation}\label{eq:total_mileages}
    S_i = C_y \sum_d \rho_d n_d \cdot s_i(d)
\end{equation}

Afterward, in order to get an estimate for the total $\text{CO}_{\text{2}}$ emissions $E_{\text{CO}_{\text{2}}}$ and potential savings of the presented rebalancing strategies, we utilized the average $\text{CO}_{\text{2}}$ emissions of \SI[per-mode=fraction]{102.35}{\gram\per\kilo\meter} and multiplied it with the calculated mileages.
Furthermore, for the \emph{Wait} and \emph{Hotspot} strategies, we applied the calculated changes from our simulation study (see Table \ref{tab:results}) to the \emph{Pickup} and \emph{Rebalancing} mileages, while assuming constant \emph{Ride} distances.
Here, we assumed that Monday to Thursday will experience the same change as the simulated Wednesday, and Friday to Sunday will behave similar to the simulated Saturday.

Results from the extrapolation are shown in Table \ref{tab:results_extrapolation}, where $S$ signifies the total mileage, $\Delta \textit{X}$ the delta to the \emph{Return} strategy, and $\bar{X}$ the average daily value.
It can be seen that, depending on the applied rebalancing strategy, the entire ride hailing fleet in Berlin could drive up to \SI{70000000}{\kilo\meter} less in a year (\SI{195000}{\kilo\meter} per day) and thereby save around \SI{7.25}{\tonne} of $\text{CO}_{\text{2}}$ emissions.
To put these numbers into perspective, the saved $\text{CO}_{\text{2}}$ emissions are equivalent to about \SI{1.8}{\permille} of the entire $\text{CO}_{\text{2}}$ emissions of the mobility sector (including planes, trains, and ships) in Berlin in 2021\cite{co2berlin}.
Furthermore the daily savable mileage of around \SI{195000}{\kilo\meter} amounts to around \SI{1.1}{\percent} of the individual motorized traffic in 2018\cite{MISberlin}. This equals to a scenario, in which suddenly \SI{7000} less private vehicles would be moving within Berlin.

\begin{table}
\caption{\textbf{Results of the Extrapolation}}
\label{tab:results_extrapolation}
\centering
\setlength{\tabcolsep}{1pt}
\begin{tabular}{|L{0.099\columnwidth}L{0.076\columnwidth}|R{0.23\columnwidth}R{0.23\columnwidth}R{0.23\columnwidth}|}
\hline
& & \multicolumn{1}{c}{\textbf{Return}} & \multicolumn{1}{c}{\textbf{Wait}} & \multicolumn{1}{c|}{\textbf{Hotspot}} \\
\hline
$S_p$ & [\si{\kilo\meter}] & \num{41409997} & \num{43668439} & \num{44423004} \\
$S_r$ & [\si{\kilo\meter}] & \num{130365143} & \num{130365143} & \num{130365143} \\
$S_b$ & [\si{\kilo\meter}] & \num{73478183} & \num{0} & \num{19546686} \\
$S$ & [\si{\kilo\meter}] & \num{245253323} & \num{174033582} & \num{194334832} \\
$\Delta S$ & [\si{\kilo\meter}] & \num{0} & \num{71219741} & \num{50918491} \\
$\bar{S}$ & [\si{\kilo\meter}] & \num{671927} & \num{476804} & \num{532424} \\
$\Delta \bar{S}$ & [\si{\kilo\meter}] & \num{0} & \num{195123} & \num{139503} \\
\hline
$E_{\text{CO}_{\text{2}}}$ & [\si{\kilo\gram}] & \num{25101678} & \num{17812337} & \num{19890170} \\
$\Delta E_{\text{CO}_{\text{2}}}$ & [\si{\kilo\gram}] & \num{0} & \num{7289340} & \num{5211508} \\
\hline
\end{tabular}
\end{table}

While these results seem promising, one has to keep in mind that many assumptions made may change in the future.
Especially the average $\text{CO}_{\text{2}}$ emissions are likely continue to decline as we move towards more climate-friendly fuel alternatives.
Additionally, only data from one fleet with a centralized hub was considered, due to which implementation of the rebalancing strategies had to follow a fairly paradigm.
Nonetheless, the captured results can be used as an upper threshold for potential reductions in mileage and emissions.

\section{Conclusion and Outlook}
\label{sec:conclusion}

In this paper, we introduced a novel simulation-based approach aimed at modeling and analyzing Ride-Hailing services, leveraging the Eclipse MOSAIC framework. Recognizing the complexity of such systems, we identified several crucial aspects that required modeling beyond traditional static OD-matrices, which proved insufficient to capture the dynamic nature of these services. We found that the MOSAIC framework is flexible enough allowing us to explore and investigate these intricate dynamics effectively.

Our analysis primarily focused on examining strategies pertaining to post-ride behavior, particularly in scenarios where returns to the Point-of-Business could be avoided. While the strategies we examined were deliberately kept simple and were not portrayed as optimal solutions, they nonetheless provided valuable insights into the potential benefits of such approaches. Furthermore, our investigation delved into different booking scenarios, with a particular emphasis on the concept of Dynamic Dispatching. We observed notable improvements in terms of distance traveled, as well as comparable or reduced wait times for passengers.

However, it is important to acknowledge that the successful implementation of these strategies relies heavily on a comprehensive understanding of the entire order situation, encompassing multiple fleet operators and the broader ecosystem of Ride-Hailing services. As such, future research in this area promises to deliver even more accurate and impactful solutions, potentially leading to significant savings in terms of distance traveled, emissions, and overall time efficiency.


\bibliographystyle{abbrv}
\bibliography{Leveraging_Eclipse_MOSAIC_for_Modeling_and_Analyzing_Ride-Hailing_Services}

\begin{thebibliography}{10}

\bibitem{matsim2016}
The multi-agent transport simulation {MATSim}, 2016-08.

\bibitem{MISberlin}
{Mobilit\"at} in {St\"adten} – {System} repr\"asentativer
  {Verkehrsbefragungen} ({SrV}) 2018, 2018.

\bibitem{pbefg}
Personenbef{\"o}rderungsgesetz {(PBefG)}, 2023.

\bibitem{berlinopendata}
Statistik {Taxen} und {Mietwagen}, 2023.

\bibitem{deng2022}
Y.~Deng, H.~Chen, S.~Shao, J.~Tang, J.~Pi, and A.~Gupta.
\newblock Multi-objective vehicle rebalancing for ridehailing system using a
  reinforcement learning approach.
\newblock {\em Journal of Management Science and Engineering}, 7(2):346--364,
  2022.

\bibitem{ester1996}
M.~Ester, H.-P. Kriegel, J.~Sander, and X.~Xu.
\newblock A density-based algorithm for discovering clusters in large spatial
  databases with noise.
\newblock In {\em Proceedings of the Second International Conference on
  Knowledge Discovery and Data Mining}, KDD'96, page 226–231. AAAI Press,
  1996.

\bibitem{co2berlin}
A.~für {Statistik} {Berlin}-{Brandenburg}.
\newblock {Energie}- und {$\text{CO}_\text{2}$}-{Bilanz} für {Berlin} und
  {Brandenburg}, 2021.

\bibitem{gueriau2020}
M.~Gueriau, F.~Cugurullo, R.~A. Acheampong, and I.~Dusparic.
\newblock Shared autonomous mobility on demand: A learning-based approach and
  its performance in the presence of traffic congestion.
\newblock {\em IEEE Intelligent Transportation Systems Magazine},
  12(4):208--218, 2020.

\bibitem{guo2021}
X.~Guo, N.~S. Caros, and J.~Zhao.
\newblock Robust matching-integrated vehicle rebalancing in ride-hailing system
  with uncertain demand.
\newblock {\em Transportation Research Part B: Methodological}, 150:161--189,
  2021.

\bibitem{guo2023fairnessenhancing}
X.~Guo, H.~Xu, D.~Zhuang, Y.~Zheng, and J.~Zhao.
\newblock Fairness-enhancing vehicle rebalancing in the ride-hailing system,
  2023.

\bibitem{hoerl2017}
S.~H{\"o}rl.
\newblock Agent-based simulation of autonomous taxi services with dynamic
  demand responses.
\newblock {\em Procedia Computer Science}, 109:899--904, 05 2017.

\bibitem{hoerl2017fleet}
S.~H{\"o}rl, C.~Ruch, F.~Becker, E.~Frazzoli, and K.~W. Axhausen.
\newblock Fleet control algorithms for automated mobility. a simulation
  assessment for zurich.
\newblock {\em Arbeitsberichte Verkehrs- und Raumplanung}, 1270, 2017-08.

\bibitem{huebner2015}
K.~H{\"u}bner, B.~Sch{\"u}nemann, and I.~Radusch.
\newblock {Sophisticated Route Calculation Approaches for Microscopic Traffic
  Simulations}.
\newblock {\em EAI Endorsed Transactions on Smart Cities}, 1(2), 8 2015.

\bibitem{emissions2015}
D.~Krajzewicz, M.~Behrisch, P.~Wagner, R.~Luz, and M.~Krumnow.
\newblock Second generation of pollutant emission models for {SUMO}.
\newblock In M.~Behrisch and M.~Weber, editors, {\em Modeling Mobility with
  Open Data}, pages 203--221, Cham, 2015. Springer International Publishing.

\bibitem{yuan2024}
Y.~Liang.
\newblock Fairness-aware dynamic ride-hailing matching based on reinforcement
  learning.
\newblock {\em Electronics}, 13(4), 2024.

\bibitem{sumo2018}
P.~A. Lopez, M.~Behrisch, L.~Bieker-Walz, J.~Erdmann, Y.~Fl{\"o}tter{\"o}d,
  R.~Hilbrich, L.~L{\"u}cken, J.~Rummel, P.~Wagner, and E.~Wie{\ss}ner.
\newblock Microscopic traffic simulation using {SUMO}.
\newblock In {\em The 21st IEEE International Conference on Intelligent
  Transportation Systems}. IEEE, 2018.

\bibitem{makhdomi2024greedy}
A.~A. Makhdomi and I.~A. Gillani.
\newblock A greedy approach for increased vehicle utilization in ridesharing
  networks, 2024.

\bibitem{taxiNyc}
{NYC Taxi and Limousine Commission}.
\newblock {TLC} trip record data, 2023.

\bibitem{protzmann2022}
R.~Protzmann, K.~Schrab, M.~Schweppenh{\"a}user, and I.~Radusch.
\newblock Implementation of a perception module for smart mobility applications
  in {Eclipse MOSAIC}.
\newblock {\em SUMO Conference Proceedings}, 3:199–214, Sep. 2022.

\bibitem{schrab2022tits}
K.~Schrab, R.~Neubauer, M.~Protzmann, I.~Radusch, S.~Manganiaris, P.~Lytrivis,
  and A.~J. Amditis.
\newblock Modeling an {ITS} management solution for mixed highway traffic with
  {Eclipse MOSAIC}.
\newblock {\em IEEE Transactions on Intelligent Transportation Systems},
  24(6):6575--6585, 2023.

\bibitem{schrab2023best}
K.~Schrab, R.~Protzmann, and I.~Radusch.
\newblock A large-scale traffic scenario of berlin for evaluating smart
  mobility applications.
\newblock In E.~G. Nathanail, N.~Gavanas, and G.~Adamos, editors, {\em Smart
  Energy for Smart Transport}, pages 276--287, Cham, 2023. Springer Nature
  Switzerland.

\bibitem{tirachini2020ride}
A.~Tirachini.
\newblock Ride-hailing, travel behaviour and sustainable mobility: an
  international review.
\newblock {\em Transportation}, 47(4):2011--2047, 2020.

\bibitem{tuncel2023}
K.~Tuncel, H.~N. Koutsopoulos, and Z.~Ma.
\newblock An integrated ride-matching and vehicle-rebalancing model for shared
  mobility on-demand services.
\newblock {\em Computers \& Operations Research}, 159:106317, 2023.

\bibitem{hbefa}
K.~Weller and S.~Hausberger.
\newblock {\em Measurement Data in the {HBEFA} database}.
\newblock 2020.

\bibitem{zwick2020}
F.~Zwick and K.~W. Axhausen.
\newblock Analysis of ridepooling strategies with {MATSim}.
\newblock Zurich, 2020-05. IVT, ETH Zurich.
\newblock 20th Swiss Transport Research Conference (STRC 2020) (virtual);
  Conference Date: May 13-14, 2020.

\end{thebibliography}

\end{document}